\documentclass[aps,prl,twocolumn,groupedaddress]{revtex4}

\begin{document}

\title{
Can $ (dG/dt) / G $ Bound the Local Cosmological Dynamics?
}

\author{Yurii V. Dumin}
\email[]{dumin@yahoo.com}
\affiliation{
Theoretical Department, IZMIRAN, Russian Academy of Sciences,
Troitsk, Moscow reg., 142190 Russia
}

\date{October 9, 2006}

\begin{abstract}
It is argued that constraints on time variation of the gravitational
constant $ (dG/dt) / G $, e.g.\ derived from the lunar laser ranging,
cannot be immediately applied to restrict the cosmological expansion at
planetary scales, as it was done by Williams, Turyshev, and Boggs
[PRL, {\bf 93,} 261101 (2004)].
\end{abstract}


\maketitle

Paper~\cite{wil04} presented the latest results
of the analysis of lunar laser ranging (LLR), one of which is
a constraint on time variation in the gravitational constant as strong as
$ \, {\dot G} / G = \! (4{\pm}9){\times}10^{-13} $~yr${}^{-1}$.
So, as was concluded by these authors,
``the $ {\dot G} / G $ uncertainty is 83~times smaller than
the inverse age of the Universe, $ t_0 = 13.4 $~Gyr~\dots \,
Any isotropic expansion of the Earth's orbit which conserves
angular momentum will mimic the effect of $ \dot G $
on the Earth's semimajor axis,
$ {\dot a} / a = - {\dot G} / G $~\dots \,
There is no evidence for such local (${\sim}1$~AU) scale expansion
of the solar system.''

Unfortunately, the above-stated equivalence between the effect of
variable $G$ and the cosmological expansion is based solely on
the Newtonian arguments. A more accurate treatment of this problem
in the framework of General Relativity for a general case of
the multi-component Universe is very difficult. Nevertheless,
for the particular case of a point-like mass~$M$ in the Universe
filled only with $\Lambda$-term (or the so-called `dark energy',
which is now assumed to be the main ingredient responsible for
the cosmological dynamics), the corresponding solution was
obtained very long time ago by Kottler~\cite{kot18}:
\begin{eqnarray}
{\rm d} s^2 \! & \!\! = \! & -\, {\Bigl( 1 - \frac{2 G M}{c^2 r'}
  - \frac{\Lambda {r'}^2}{3} \Bigr)}\, c^2 {\rm d}{t'}^2
\label{eq:Kottler_metric}
\\
  & & + \, {\Bigl( 1 - \frac{2 G M}{c^2 r'}
  - \frac{\Lambda {r'}^2}{3} \Bigr)}^{\!\! -1} \!\! {\rm d}{r'}^2 \!
  + \: {r'}^2 ( {\rm d}{\theta}^2 \!\!\!
  + {\sin}^2{\theta} \, {\rm d}{\varphi}^2 ) \, ,
\nonumber
\end{eqnarray}
where $G$ is the gravitational constant,
and $c$ is the speed of light.
(For a general review, see also~\citep{kra80}.)

After a transformation to the cosmological Robertson--Walker coordinates
\begin{eqnarray}
r' \!\!\! & = & a_0 \, {\exp} \Bigl( \frac{\displaystyle c t}%
     {\displaystyle r_0} \Bigr) \, r \, ,
\label{eq:r_coord}
\\
t' \!\!\! & = &
     t - \frac{1}{2} \, \frac{\displaystyle r_0}{\displaystyle c} \:\:
     {\ln} \Bigl[ 1 - \frac{\displaystyle a_0^2}{\displaystyle r_0^2} \:
     {\exp} \Bigl( \frac{\displaystyle 2 c t}{\displaystyle r_0} \Bigr)
     \: r^2 \Bigr] \, ,
\label{eq:t_coord}
\end{eqnarray}
metric~(\ref{eq:Kottler_metric}) takes the form
\begin{eqnarray}
{\rm d} s^2 \!\! & \!\! = \! & \! g_{tt} \, c^2 {\rm d}{t}^2
  + \, 2 \, g_{tr} \, c \, {\rm d}{t} \, {\rm d}{r}
  + \, g_{rr} \, {\rm d}{r}^2
\nonumber
\\
  & \! + & \! g_{\theta \theta} \, {\rm d}{\theta}^2
  + \,g_{\varphi \varphi} \, {\rm d}{\varphi}^2 \, ,
\label{eq:metric_cosm_coord}
\end{eqnarray}
where
\begin{eqnarray}
g_{tt} \!\! & = &
  \frac{\displaystyle - {\Bigl( 1 - \frac{r_g}{r'} - \frac{{r'}^2}{r_0^2}
\Bigr)}^{\!\! 2} + {\Bigl( 1 - \frac{{r'}^2}{r_0^2} \Bigr)}^{\!\! 2}
\frac{{r'}^2}{r_0^2}}%
       {\displaystyle {\Bigl( 1 - \frac{r_g}{r'} - \frac{{r'}^2}{r_0^2}
\Bigr)} {\Bigl( 1 - \frac{{r'}^2}{r_0^2} \Bigr)}^{\!\! 2} } \:\: ,
\label{eq:g_tt}
\\
g_{tr} \!\! & = &
  \frac{\displaystyle {\Bigl( 1 - \frac{{r'}^2}{r_0^2} \Bigr)}^{\!\! 2}
- {\Bigl( 1 - \frac{r_g}{r'} - \frac{{r'}^2}{r_0^2} \Bigr)}^{\!\! 2}}%
       {\displaystyle {\Bigl( 1 - \frac{r_g}{r'} - \frac{{r'}^2}{r_0^2}
\Bigr)} {\Bigl( 1 - \frac{{r'}^2}{r_0^2} \Bigr)}^{\!\! 2}}
  \; \frac{r'}{r_0} \, \frac{r'}{r} \:\: ,
\label{eq:g_tr}
\\
g_{rr} \!\! & = &
  \frac{\displaystyle {\Bigl( 1 - \frac{{r'}^2}{r_0^2} \Bigr)}^{\!\! 2}
- {\Bigl( 1 - \frac{r_g}{r'} - \frac{{r'}^2}{r_0^2} \Bigr)}^{\!\! 2}
\frac{{r'}^2}{r_0^2} }%
       {\displaystyle {\Bigl( 1 - \frac{r_g}{r'} - \frac{{r'}^2}{r_0^2}
\Bigr)} {\Bigl( 1 - \frac{{r'}^2}{r_0^2} \Bigr)}^{\!\! 2}}
  \; \frac{{r'}^2}{r^2} \:\: ,
\label{eq:g_rr}
\\
g_{\theta \theta} \!\! & = & \!\!
  \, g_{\varphi \varphi} / {\sin}^2 {\theta} = {r'}^2 \! .
\label{eq:g_angl}
\end{eqnarray}
In the above formulas,
$ r_g \! = 2 G M / c^2 $, $ r_0 = \sqrt{ 3 / \Lambda } \, $, and
$a_0$~is the scale factor of FRW Universe.

Taking $ a_0 \! = 1 $ at $ t \! = 0 $ and keeping only
the lowest-order terms of $ \, r_g $ and $ 1 / r_0 $, we get
\begin{eqnarray}
g_{tt} \!\! & \approx &
  - \Bigl[ 1 - \frac{2 G M}{c^2 r}
  \Bigl( 1 - \frac{c \sqrt{\Lambda} \, t}{\sqrt{3}} \Bigr) \Bigr] \, ,
\label{eq:g_tt_approx}
\\
g_{tr} \!\! & \approx &
  \frac{4 \, G M \sqrt{\Lambda}}{\sqrt{3} \, c^2} \:\: ,
\label{eq:g_tr_approx}
\\
g_{rr} \!\! & \approx &
  \Bigl[ 1 + \frac{2 G M}{c^2 r}
  \Bigl( 1 - \frac{c \sqrt{\Lambda} \, t}{\sqrt{3}} \Bigr) \Bigr]
  \Bigl( 1 + \frac{2 c \sqrt{\Lambda} \, t}{\sqrt{3}} \Bigr) \, ,
\label{eq:g_rr_approx}
\\
g_{\theta \theta} \!\! & = & \!
g_{\varphi \varphi} / {\sin}^2 {\theta} \approx
  \, r^2 \Bigl( 1 + \frac{2 c \sqrt{\Lambda} \, t}{\sqrt{3}} \Bigr) \, .
\label{eq:g_angl_approx}
\end{eqnarray}

As is seen from the above expressions, manifestation of
the $\Lambda$-term in some components of the metric tensor
really looks like the influence of variable~$G$ if we assume that
$ G = G_0 + {\dot G} \, t $, where
$ {\dot G} = - c \sqrt{\Lambda / 3} \: $.
Unfortunately, such interpretation is not self-consistent:
the $\Lambda$-dependence of a few other components is irreducible
to the variable coefficient of gravitational coupling.

Therefore, the available limits on $ \, {\dot G} / G \, $,
in general, cannot be reinterpreted as a constraint on
the local cosmological dynamics.


\end{document}